# Thin-film Lithium Niobate Dual-Polarization IQ modulator on a Silicon Substrate for Single-Carrier 1.6 Tb/s Transmission

XUANHAO WANG,[1] CHENGLIN SHANG,[1] AN PAN,[1] XINGRAN CHENG,[1] TAO GUI,[2] SHUAI YUAN,[2] CHENGCHENG GUI,[3] KESHUANG ZHENG,[2] PEIJIE ZHANG,[3] XIAOLU SONG,[3] YANBO LI,[2] LIANGCHUAN LI,[2] CHENG ZENG,[1,4] JINSONG XIA[1,5]

[1]*Wuhan National Laboratory for Optoelectronics, School of Optical and Electronic Information, Huazhong University of Science and Technology, Wuhan 430074, China*
[2]*Huawei Technologies, Optical R&D Dept., Dongguan 523000, China*
[3]*Huawei Technologies, B&P Lab., Shenzhen 518000, China*
[4]*e-mail: zengchengwuli@hust.edu.cn*
[5]*e-mail: jsxia@hust.edu.cn*

**We successfully demonstrate a monolithic integrated dual-polarization (DP) IQ modulator based on thin-film lithium niobate (TFLN) platform with a silicon substrate, which consists of IQ modulators, spot-size converters (SSCs) and a polarization rotator combiner (PRC). After coupled with polarization maintaining fibers, the measured insertion loss of the modulator is 12 dB. In addition, we experimentally achieve a single-carrier 1.6 Tb/s net bitrate transmission.** 

## 1. Introduction

With the explosive growth of internet data traffic, the demand for optical interconnection capacity is increasing in both data-center and backbone network. According to the Ethernet roadmap[1], 800 G and 1.6 T will be the next-generation optical interface speeds after 2023. Considering the limited bandwidth of optical fibers and amplifiers, high spectrum efficiency transmission has become the focus of attention. Coherent transmission technology has been proven to effectively improve spectrum efficiency utilizing advanced modulation format as well as polarization division multiplexing (PDM)[2, 3]. In-phase/quadrature (IQ) modulator is a core device for coherent transmission system, which allows encoding of information onto both the amplitude and phase of light. For decades, enormous efforts have been made to achieve IQ modulators on various material platforms, including the silicon (Si)[4-6], indium phosphide (InP)[7-9], plasmonic[10, 11], and silicon-organic hybrid (SOH)[12, 13]. The modulators on each above platforms have their own limitations, such as high driving voltage and limited bandwidth for Si, complex and expensive monolithic integration fabrication processes for InP, large loss for plasmonic and the instability of material for SOH. It is always believed that lithium niobate (LN) is an excellent material for high performance electro-optic (EO) modulators, due to its strong linear EO effect, wide transparency bandwidth and long-term good temperature stability.

Thin-film lithium niobate (TFLN) has recently emerged as an appealing material platform, which is expected to overcome the low modulation efficiency and limited EO bandwidth in traditional bulk LN modulators[14-17]. The world first IQ modulator has been demonstrated on TFLN platform with 48 GHz EO bandwidth, supporting modulation data rate up to 320 Gb s−1[18]. The voltage-bandwidth limitation in TFLN based modulator was then broken by introducing capacitively-loaded travelling-wave electrode, showing the potential to enable sub-volt modulator with >100 GHz bandwidth[19]. Besides, a 1.58 Tb/s net rate transmission was demonstrated with a TFLN based IQ modulator and an external PDM emulator[20], which proves that TFLN based IQ modulators can be used for ultra-high-speed interconnection. To date, the dual-polarization (DP) IQ modulator composed of monolithic integrated IQ modulators, spot-size converters (SSCs) and polarization rotator combiners (PRCs) is rarely reported on TFLN platform, due to the more complex fabrication processes and low yield. An excellent voltage-bandwidth performance DP-IQ modulator was recently achieved on TFLN platform with quartz substrate[21]. Quartz substrate helps to reduce microwave attenuation and thus contributes to higher EO bandwidth. However, the production of TFLN wafer with quartz substrate suffers from higher cost, lower yield and smaller size than that with Si substrate. In addition, Si substrate may be a better candidate for heterogeneous bonding with III-V distributed feedback (DFB) lasers to achieve an integrated TFLN optical transmitter[22]. As coherent transmission technology sinks to data-center interconnect, low-cost and highly integrated optical components are demanded. Therefore, it is suitable to use conventional Si substrate wafers for practical production. Particularly, a recently reported method of partially removing the Si substrate underneath the electrode of LN modulators significantly reduces the microwave loss, which enables the EO bandwidth close to modulators with quartz substrate[23, 24].

In this paper, we demonstrate a high performance TFLN based monolithic integrated DP-IQ modulator on a silicon substrate. The proposed modulator exhibits an insertion loss of 12 dB per polarization, 6 dB electro-optic bandwidth larger than 67GHz, and polarization extinction ratio of 33 dB. Furthermore, a single-carrier probability-shaping 256 QAM transmission with a net bitrate of 1.6 Tb/s via the proposed DP-IQ modulator is successfully demonstrated.

## 2. TFLN based DP-IQ Modulator

The schematic of the proposed TFLN based DP-IQ modulator is shown in Fig. 1(a). The device is designed on the 500-nm-thick X-cut TFLN wafer. The DP-IQ modulator consists of monolithic integrated two IQ modulators, several SSCs and one PRC. The IQ modulators are driven by 1.2-cm-long ground–signal–ground-signal–ground (GSGSG) traveling-wave electrode in a push-pull mode. The coupling of light between the fiber and the LN waveguide is through the low loss and polarization-independent SSCs[25]. The 1 × 2 multimode interference (MMI) couplers are used as symmetrical optical power splitters/combiners. Six thermo-optic phase shifters (PSs) are employed to control the modulation bias points of the Mach-Zehnder modulators (MZMs), and π/2 phase shift between I and Q components, respectively. The thermo-optical PSs are formed by Ti thin film heaters, which can avoid the DC drift phenomenon in EO effect[18]. To achieve PDM, a PRC cascading an adiabatic taper and an asymmetric Y-junction is applied to combine lights from two IQ modulators into orthogonal polarization states, the design details of the PRC are shown in [26]. The microscope images of the thermo-optical PS and PRC are shown in Fig. 1(b-c).

The high-speed modulation section is the core part of the modulator, which consists of optical waveguides and coplanar waveguide (CPW) electrodes. The etching depth of LN ridge waveguide is 260 nm, and 70°-waveguide sidewall angle is caused by the fabrication processes. The cross-section view of the modulation section is shown in Fig. 2(a). The LN ridge waveguide has a width of W = 1.5 μm in the modulation section. The electrode gap is set to G=5 μm by a Finite Difference Eigenmode (FDE) solver, to achieve the maximum modulation efficiency while ensuring the electrodes will not introduce additional optical absorption loss. The calculated mode field distribution of the TE0 mode in LN waveguide is shown in Fig. 2(b). Then the CPW electrode is designed for velocity matching of the microwave and optical signals, as well as impedance matching. We optimize The CPW electrode structure by Finite Element Method (FEM). The width of signal and ground electrodes are designed as $W_S$ = 40 μm and $W_G$ = 50 μm, respectively. The thickness of the Au electrode is H = 1200 nm. A 2-μm-thick $SiO_2$ layer covers the electrode as the upper cladding. Fig. 2(c) shows the calculated electric field distribution when a 1-V voltage is applied, thus we can get the theoretical voltage-length product of 2.43 V·cm. The RF effective index is 2.3 matched to the optical group index, and the characteristic impedance of the CPW electrode is 38 Ω. The electrode impedance is less than 50 Ω as a compromise between characteristic impedance and microwave loss. Narrower signal electrode could be used to increase characteristic impedance. However, this will increase microwave loss and further degrade the EO bandwidth. The EO response is also simulated as shown in Fig. 2(d). We get the 3 dB bandwidth of 51 GHz when the termination impedance is matched.

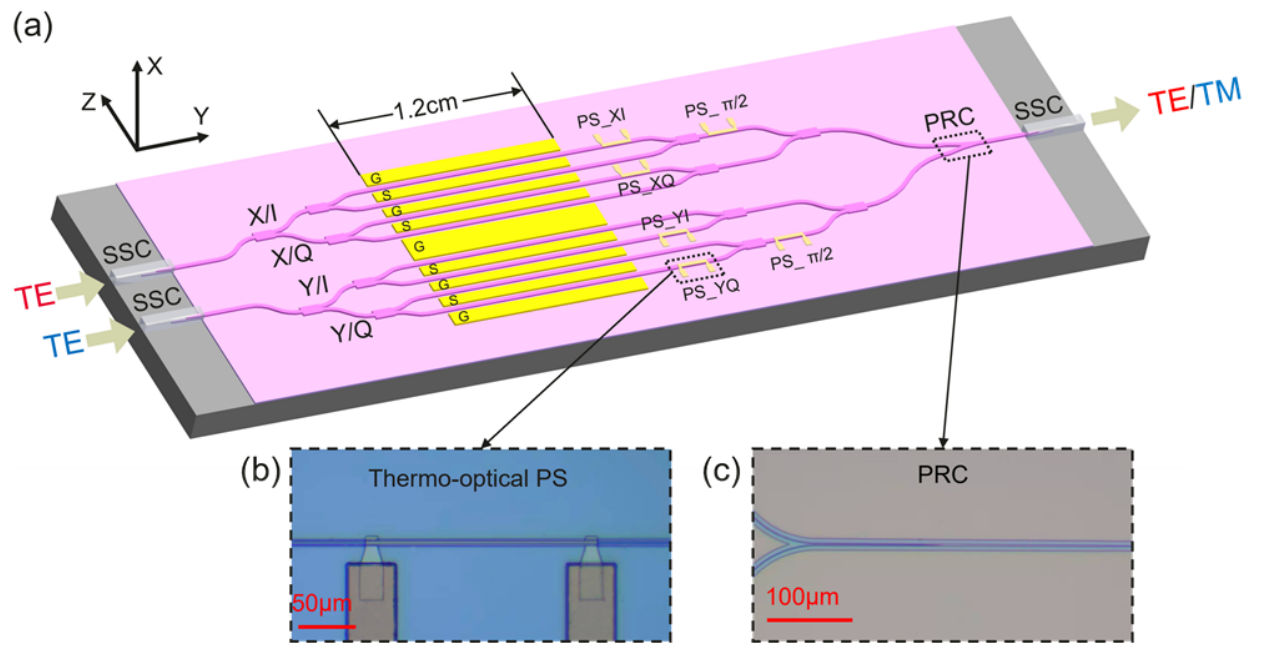

**Fig. 1.** (a) Schematic structure of the DP-IQ modulator. (b) Microscope image of the thermo-optical PS. (c) Microscope image of the PRC.

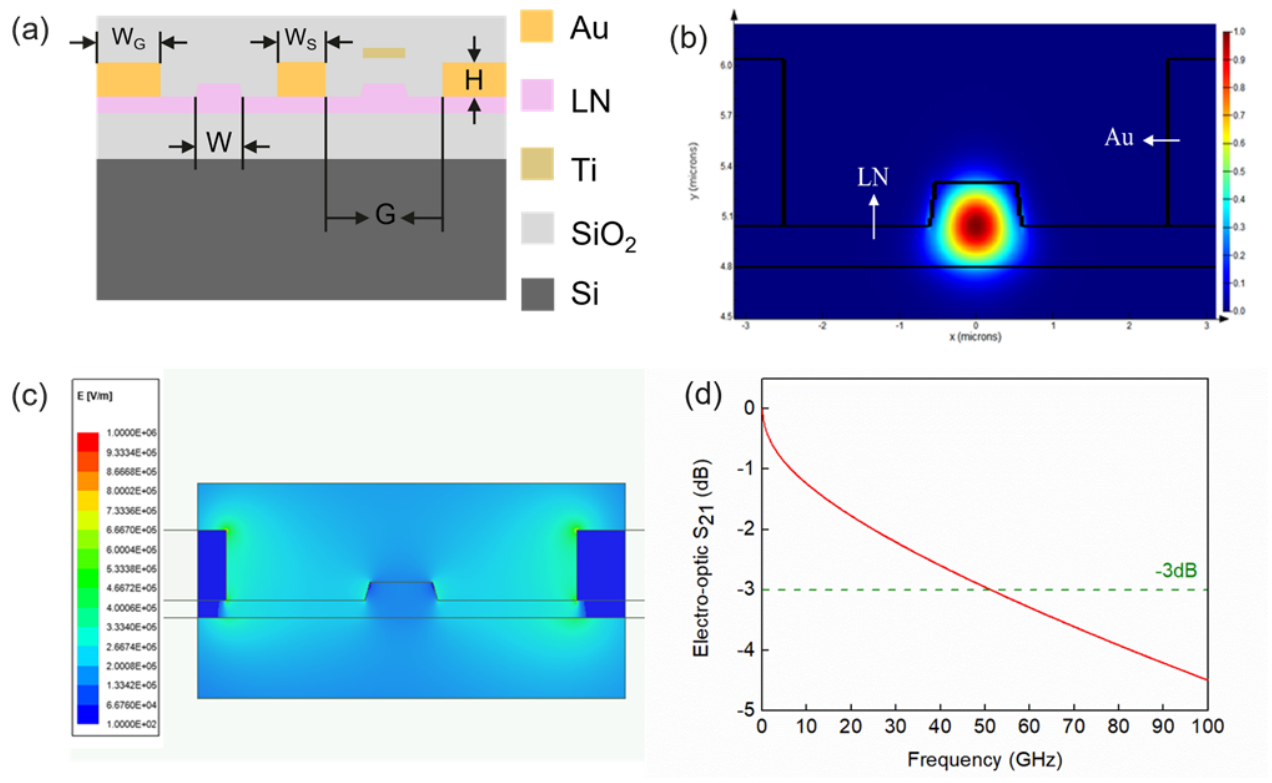

**Fig. 2.** (a) Cross-section view of the high-speed modulation section. (b) The TE0 mode field distribution in LN waveguide. (c) The electric field distribution when a voltage of 1 V is applied. (d) The simulated electro-optic $S_{21}$ of the modulator

The DP-IQ modulator is fabricated on a commercial lithium niobate on insulator (LNOI) wafer (NANOLN), which includes 500-µm silicon substrate, 4.7-µm buried oxide and 500-nm LN thin film. Electron beam lithography (EBL), inductively coupled plasma reactive-ion-etching (ICP-RIE), electron-beam evaporation (EBE), plasma enhanced chemical vapor deposition (PECVD) and end face polishing processes are used during the fabrication. After the fabrication of the TFLN based DP-IQ modulator chip, the optical coupling between the LN waveguide and the polarization maintaining fiber array is achieved by UV-curable glue, as shown in Fig. 3(a). We use polarization maintaining fiber to ensure TE mode coupling into the LN waveguide. An external termination resistor (38 Ω) is placed at the end of the electrode to achieve impedance match. The modulator chip is then mounted on a printed circuit board (PCB), and the electrode pads of PSs are connected to the PCB pads through gold wires. To measure the fiber-to-fiber insertion loss and polarization extinction ratio, a polarization controller (PC) is put behind the tunable laser (Santec TLS-510) to control the polarization state of light. The light is coupled into the chip from the output port of the modulator. The measured fiber to fiber insertion loss of the simple packaged DP-IQ modulator is 12 dB per polarization, including 6 dB from two SSCs and on-chip loss of 6 dB. On the other hand, we get the polarization extinction ratio over 33 dB at 1550 nm by comparing the transmittance of TE and TM modes. The coupling loss of SSC is evaluated to be 3 dB/facet via a straight waveguide with SSC at both ends on the same chip. Compared to our previous results[25], the coupling loss of SSC increases significantly. The coupling loss is dominated by the mode field mismatch between polarization maintaining fiber (mode field diameter is 6.5 µm) and SSC (mode field diameter is 3.2 µm). Meanwhile, the position shift of the fiber array caused by UV-curable glue during the 80 ℃ thermal cure also affects the coupling loss. The on-chip loss is mainly caused by the propagation loss of total 6.3-cm-long waveguide. We will further optimize the chip fabrication and optical packaging processes to reduce the insertion loss of the modulator.

When measuring the half-wave voltage of the modulator, a 1 kHz triangle electric signal is applied onto the MZMs. As shown in Fig. 3(b), the half-wave voltage is 2.18 V, corresponding to the half-wave voltage length product of 2.6 V·cm. The extinction ratios of the MZMs are around 24 dB. The small-signal EO response of the modulator is also characterized. The RF signal is added to the GSGSG electrodes via two 67-GHz RF probes. The measured S21 and electrical reflections (S11) of the XI, XQ, YI and YQ MZMs are shown in Fig. 3(c), where the dashed line shows the simulated S21 curve. It is worth mentioning that the S11 and S21 curves are obtained without excluding the effect of RF probes. We observe a 3 dB bandwidth roll-off around 30 GHz, and S11 of all MZMs are less than -14 dB. The measured S21 curve drops slightly faster than simulated results, due to the RF attenuation caused by the probes. However, it is possible for the modulator to operate at a frequency far beyond the 3dB bandwidth considering the flatness of the S21 curve. We can observe a 6dB electro-optical bandwidth greater than 67GHz. We successfully demonstrate a TFLN based DP-IQ modulator, which is more complex in fabrication than previously reported single polarization TFLN modulator.

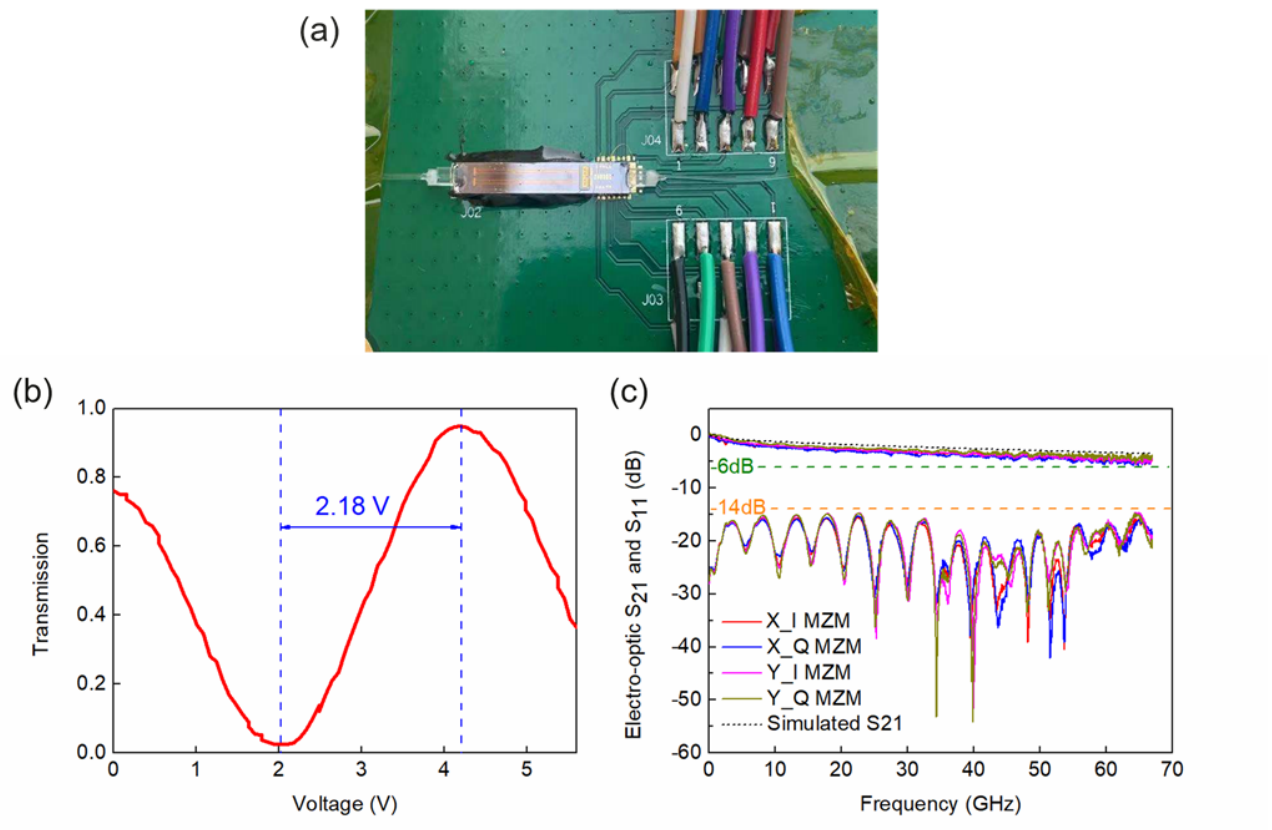

**Fig. 3.** (a) Image of the simple packaged DP-IQ modulator chip. (b) Normalized optical transmission of MZM as a function of the applied voltage. (c) Electro-optic responses and electrical reflections (S11) of the 4 channel MZMs

## 3. Transmission Experiment

We use our DP-IQ modulator to demonstrate high-speed coherent transmission. The experimental setup is shown in Fig. 4(a). 136-GBaud single carrier probability-shaping DP-256QAM signal is generated by an arbitrary-waveform generator (AWG, Keysight 8199A) with a typical bandwidth of 70 GHz and sampling rate of 224 GSa/s, and the entropy is about 7.0 bits/symbol[27]. Open FEC with 15% overhead is assumed, and 3.5% overhead pilot is used[28]. The data signal output from the DAC consists of four real components, which are connected to four single-ended RF amplifiers (SHF S804 B) with 60 GHz 3 dB-bandwidth and 22 dB gain to drive our DP-IQ modulators. The optical spectra are shown as inset to Fig. 4(a). The modulated signal is transmitted through a 2-km-long single-mode fiber, and then down-converted to the electrical domain using four balanced PDs (6 dB attenuation at 100 GHz) at the coherent receiver. The down converted signal is sampled by a 256-GSa/s 110-GHz real-time scope followed by offline DSP. In the transmitter DSP, after mapping and framing, pulse shaping with 0.01 roll-off factor, nonlinear and linear pre-distortion are performed. At the receiver, the DSP steps are: i) resampling; ii) frame synchronization; iii) a real-valued 4 × 2 least mean square (LMS) equalizer; iv) down-sampling; v) carrier phase recovery aided by training and pilot symbols; vi) 2 × 2 LMS equalizers; vii) whitening filter; viii) 2 × 2 decision feedback equalizer; ix) symbol decision.

The measured BER performances as a function of OSNR is shown in Fig. 4(b). The pre-FEC BER threshold is 2.0E-2. When OSNR is 36 dB, the BERs are below the pre-FEC threshold, which indicates error free will be achieved after

FEC decoding. The corresponding recovered probability-shaping 256QAM constellations are shown in Fig. 4(c). We successfully demonstrated a single-carrier 1.6 Tb/s net bitrate transmission via our TFLN based DP-IQ modulator.

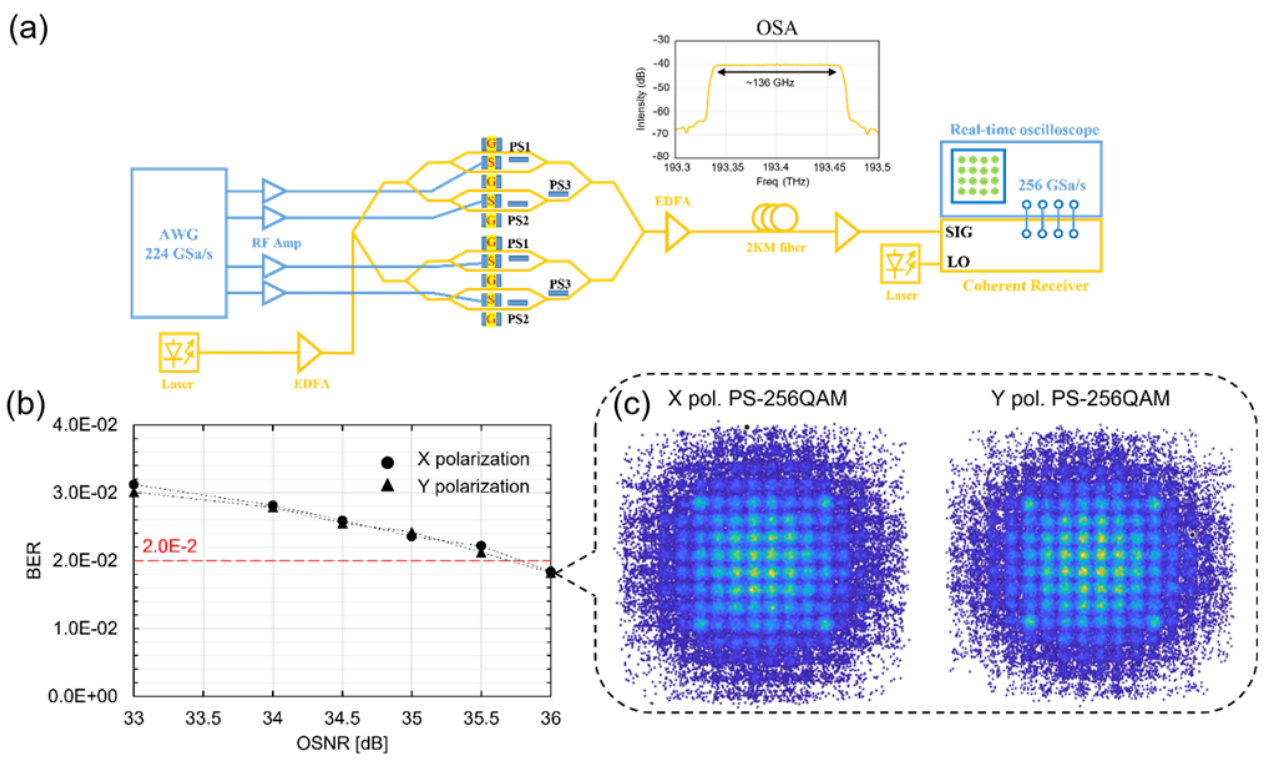

**Fig 4.** (a) Schematic of the experimental setup for coherent data transmission, inset: optical spectra of the transmitted signal. (b) Measured BER performances after transmission. (c) recovered constellation diagrams for probability-shaping DP-256QAM signal.

## 4. Conclusions

A monolithic integrated DP-IQ modulator is realized on the TFLN platform with a silicon substrate, which consists of IQ modulators, on-chip SSCs and a PRC. After simple packaging, the proposed modulator exhibits an insertion loss of 12 dB, 6 dB electro-optic bandwidth larger than 67GHz (including the effect of RF probe), and polarization extinction ratio of 33 dB. Furthermore, a single-carrier probability-shaping DP-256QAM Transmission with a net bitrate of 1.6Tb/s via the proposed DP-IQ modulator is successfully achieved. We believe this is a promising solution for future high speed optical interconnects.

**Funding.** This work is supported by National Key Research and Development Program of China (2021YFB2800104, 2019YFB2203501), and the National Natural Science Foundation of China under Grant No. 61835008, 62175079, 61905079, 61905084.

**Acknowledgments.** We thank the Center of Micro-Fabrication and Characterization (CMFC) of WNLO and the Center for Nanoscale Characterization & Devices (CNCD), WNLO of HUST for the facility support.

**Disclosures.** The authors declare no conflicts of interest.

**Data availability.** Data underlying the results presented in this paper are not publicly available at this time but may be obtained from the authors upon reasonable request.